\documentclass[onecollarge,natbib]{svjour2}
\bibpunct{[}{]}{;}{n}{}{,} 
\smartqed  
\usepackage{graphicx}
\usepackage{amssymb,amsmath,epsfig,bm,pifont}
\usepackage{relsize}
\newcommand{\babar}{{\mbox{\slshape B\kern-0.1em{\smaller A}\kern-0.1em
            B\kern-0.1em{\smaller A\kern-0.2em R}}}
           }
\journalname{Few-Body Systems}
\begin{document}

\title{Pion-photon transition form factor.\thanks{Presented at LIGHTCONE 2011, 23 - 27 May, 2011, Dallas, USA}
      }
\subtitle{Living on the QCD frontier}

\titlerunning{Pion-photon transition}        

\author{N.~G.~Stefanis
       }


\institute{N.~G.~Stefanis \at
           Institut f\"{u}r Theoretische Physik II,
           Ruhr-Universit\"{a}t Bochum, D-44780 Bochum, Germany \\
           Tel.: +49-234-3223724 \\
           Fax:  +49-234-3214248 \\
           \email{stefanis@tp2.ruhr-uni-bochum.de}           
}

\date{Received: 31-08-2011 / Accepted: 12-09-2011}

\maketitle

\begin{abstract}
An analysis of all available data (CELLO, CLEO, \babar$\!\!$)
in the range $[1\div 40]$~GeV$^2$ for the pion-photon transition
form factor in terms of light-cone sum rules with
next-to-leading-order accuracy is discussed, including twist-four
contributions and next-to-next-to-leading order and twist-six
corrections---the latter two via uncertainties.
The antithetic trend between the \babar data for the
$\gamma^*\gamma\pi^0$ and those for the $\gamma^*\gamma\eta(\eta')$
transition is pointed out, emphasizing the underlying antagonistic
mechanisms: endpoint enhancement for the first and
endpoint-suppression for the second---each associated with pseudoscalar
meson distribution amplitudes with distinct endpoint characteristics.
\end{abstract}

\section{Introduction}
\label{intro}
This is a report about the pion-photon transition form factor---an
observable that is considered to be the prime example of an exclusive
process in QCD because in leading order (LO) the partonic interactions
are purely electrodynamic, with the strong interactions being factorized
out into the pion distribution amplitude (DA).
While the accuracy of the CELLO data \cite{CELLO91} was not sufficient
to provide a proof-of-concept of the QCD framework, the appearance of
the CLEO data \cite{CLEO98} established the validity of the collinear
factorization of this process \cite{LB80}.
This situation changed dramatically in the year 2009 when the \babar
Collaboration published new data \cite{BaBar09} with very small error
bars at intermediate momentum transfers reaching values just
below $40$~GeV$^2$---unreachable in the past---with still tolerable
errors.
These data for the scaled form factor
$Q^2F^{\gamma^*\gamma\pi^0}(Q^2)$
exhibit two distinctive features:
(i) The earlier data taken by CLEO at the \emph{same}
momentum-transfer values were confirmed, but with a higher accuracy.
(ii) Unexpectedly, the data starting above $9$~GeV$^2$ and extending up
to about $40$~GeV$^2$ do not scale with $Q^2$.
Instead, they show a clear general tendency to rapidly grow with $Q^2$,
with the exception of two single data points at about 14~GeV$^2$ and
27~GeV$^2$ that lie just below the asymptotic prediction of
perturbative QCD \cite{LB80}:
$
 \lim_{Q^2\to\infty} Q^2F^{\gamma^*\gamma\pi^0}
 \longrightarrow
 \sqrt{2}f_\pi
$.
Understanding the underlying enhancement mechanism(s), responsible
for this behavior, is of significant importance because an increasing
form factor challenges the basic concepts of QCD, like collinear
factorization.
The recent analysis in Ref.\ \cite{BMPS11}, reported here, builds on
light-cone sum rules (LCSR)s by taking into account various
contributions from perturbative and nonperturbative QCD and by
utilizing in the analysis data sets from different experiments
\cite{CELLO91,CLEO98,BaBar09} for momentum-transfer values ranging
from 1 to 40~GeV$^2$.
Emphasis is given to the statistical features and the way they depend
on the various theoretical parameters.
The main goal is the confrontation of theoretical predictions,
obtained with the help of light-cone sum rules (LCSR)s, with all
available experimental data for the pion-photon transition form factor.
In addition, the results are compared with the \babar data
\cite{BaBar11-BMS} for the $\eta$- and $\eta^\prime$-photon transition
form factors making use of the description of the $\eta-\eta^{\prime}$
mixing in the quark flavor basis \cite{FKS98} that involves the state
$|n\rangle = (|\bar{u}u\rangle + |\bar{d}d\rangle)/\sqrt{2}$.
This makes it possible to link the
$\gamma^*\gamma\to |n\rangle$ transition form factor,
multiplied by $3/5$, to the form factor $\gamma^*\gamma\to\pi^0$,
where the prefactor arises from the quark charges.

\section{Computational Method}
\label{sec:method}
The method of light-cone sum rules, developed in \cite{Kho99}
for the pion-photon transition form factor, but not only \cite{BBK89},
has been refined by years of experience in many analyses
\cite{SY99,BMS02-05,MS09,ABOP10,BMPS11}.
It contains a set of rules that encode a computational model and as any
model it encodes \emph{implicit} predictions that turn \emph{explicit}
under some justified assumptions about its inherent (and auxiliary)
parameters.
The subtlety is how to adjust these parameters in such a way as to
achieve the best possible accuracy of the obtained predictions with
respect to the experimental data while avoiding sensitivity on
particular values.

The modus operandi of the LCSRs for the pion-photon transition form
factor is expressed by \cite{Kho99}
\begin{equation}
  Q^2 F^{\gamma^*\gamma\pi}\left(Q^2\right)
= \frac{\sqrt{2}}{3}f_\pi
    \left[\frac{Q^2}{m_{\rho}^2} \int_{x_{0}}^{1}
           \exp\left(\frac{m_{\rho}^2-Q^2\bar{x}/x}{M^2}
               \right)
             \bar{\rho}(Q^2,x) \frac{dx}{x}
    + \int_{0}^{x_0} \bar{\rho}(Q^2,x) \frac{dx}{\bar{x}}
    \right] \, ,
\label{eq:LCSR-FF}
\end{equation}
where the abbreviations
$s =\bar{x}Q^2/x$ and
$x_0 = Q^2/(Q^2+s_0)$
have been used, and with the spectral density
$\bar{\rho}(Q^2,x)=(Q^2+s)\rho^{\rm PT}(Q^2,s)$ taken with
next-to-leading-order (NLO) accuracy.
The hadronic threshold in the vector-meson channel has the value
$s_0=1.5$~GeV$^2$, $M$ is the Borel parameter, and $m_{\rho}=0.77$~GeV
denotes the physical mass of the $\rho$ meson.
The first term in (\ref{eq:LCSR-FF}) stems from the hadronic content
of a quasireal photon at low $s\leq s_0$, while the second one
resembles its pointlike behavior at higher $s>s_0$.
The main advantage of employing LCSRs is that one starts with the
situation where both photon virtualities are sufficiently large, so
that perturbation theory is safely applicable, and approaches the
asymmetric kinematics with $Q^2$ fixed and large and $q^2\to 0$
via a dispersion relation:
\begin{equation}
  F^{\gamma^{*}\gamma^{*}\pi}\left(Q^2,q^2\right)
= \int_{0}^{\infty}\!\!ds\,
  \frac{\rho\left(Q^2,s\right)}{s+q^2}\, ,
\label{eq:dis-rel}
\end{equation}
where the physical spectral density $\rho(Q^2,s)$
approaches at large $s$ the perturbative one:
$
 \rho^{\rm PT}(Q^2,s)
=
 \frac{1}{\pi} {\rm Im}F^{\gamma^*\gamma^*\pi}
 \left(Q^2,-s-i\varepsilon\right)\, .
$
The form factor $F^{\gamma^{*}\gamma^{*}\pi}\left(Q^2,q^2\right)$
can be calculated via perturbative QCD having recourse to collinear
factorization, meaning that it can be written in the convolution form
\cite{LB80,ER80tmf}
\begin{equation}
  F^{\gamma^{*}\gamma^{*}\pi}(Q^2,q^2)
=
  \frac{\sqrt{2}}{3}f_\pi\int_{0}^{1}\!dx\,
  T(Q^2,q^2,\mu^2_{\rm F},x)
  \varphi^{(2)}_{\pi}(x,\mu^2_{\rm F})
  + O\left(\delta^2/Q^{4}\right)\, ,
\label{eq:convolution}
\end{equation}
where the pion decay constant is $f_\pi=132$~MeV and $\delta^2$ is
the twist-four coupling.
All nonperturbative information of the bound state is encapsulated in
the coefficients $a_n$ that are incalculable within perturbation
theory and have to be modeled, or be computed on the lattice, e.g.,
\cite{Lat06,Lat10}, via the moments
$
\langle \xi^{N} \rangle_{\pi}
\equiv
  \int_{0}^{1} dx (2x-1)^{N} \varphi_{\pi}^{(2)}(x,\mu^2)
 $
with the normalization condition
$\int_{0}^{1} dx \varphi_{\pi}^{(2)}(x, \mu^2)=1$.
The leading twist-two pion DA fulfills an evolution equation
\cite{LB80,ER80tmf} and can be expressed in terms of the Gegenbauer
polynomials $C^{3/2}_{n}(2x-1)$ to read
\begin{eqnarray}
  \varphi^{(2)}(x, \mu^2)
= \varphi^{\rm as}(x)
  + \sum\nolimits_{n=2,4,\ldots}
  a_{n}\left(\mu^2\right) \psi_{n}(x)\,,
\label{eq:pion-DA.Geg}
\end{eqnarray}
where $\psi_{n}(x)\!=\!6x(1-x) C^{3/2}_{n}(2x-1)$
and
$\varphi^{\rm as}(x)\!=\!\psi_{0}(x)\!=\!6x(1-x)$ is the asymptotic
pion DA~\cite{LB80,ER80tmf}.

\begin{figure}
\centering
\includegraphics[width=0.50\textwidth]{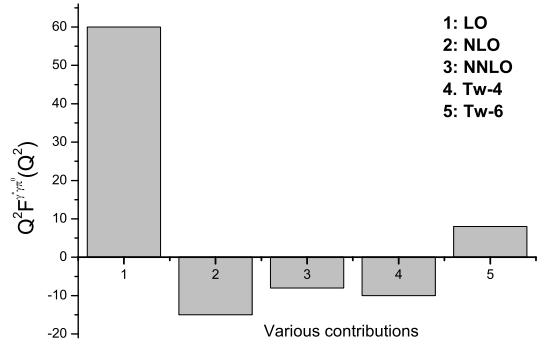}
\caption{(color online) Illustration of the various contributions to the
pion-photon transition form factor.}
\label{fig:1}
\end{figure}

The cornerstones of the analysis in \cite{BMPS11}, on which this report
is based, are:
\begin{itemize}
\item NLO radiative corrections: They are taken into account in the
spectral density \cite{MS09}, after correcting the error pointed out in
\cite{ABOP10}, and using $\mu_{\rm fact}^2=\mu_{\rm ren}^2=Q^2$
to avoid (large) logarithms of these scales.
They are negative leading to suppression of the transition form
factor.
\item Twist-four contributions: They are included via the coupling
$\delta^2(\mu^2)\in [0.15\div 0.23]$~GeV$^2$ and performing one-loop
evolution assuming the asymptotic form for the twist-four pion DA:
$
 \varphi_{\pi}^{(4)}(x,\mu^2)
=
 (80/3) \delta^2(\mu^2)\,x^2(1-x)^2
$.
[Nonasymptotic forms were found before to be of minor importance.]
These contributions are also negative and yield suppression.
\item Evolution of the coefficients $a_n$:
They are included at the NLO level of accuracy, using the QCD scale
parameters
$\Lambda_{\rm QCD}^{(N_{f}=3)}=370$~MeV and
$\Lambda_{\rm QCD}^{(N_{f}=4)}=304$~MeV,
and provide suppression.
\item NNLO$_{\beta_{0}}$ (for next-to-next-to-leading) radiative
corrections \cite{MMP02} and twist-six contributions \cite{ABOP10}:
They are taken into account implicitly in terms of theoretical
uncertainties.
This is justified because for the average value of
$M^2(Q^2)\sim 0.75$~GeV$^2$, used in our analysis \cite{BMPS11}, the
net result of adding these (not large) terms is small, decreasing with
$Q^2$ from $+0.005$ at $Q^2=1$~GeV$^2$---where the positive twist-six
term prevails---down to $-0.003$ at $Q^2=40$~GeV$^2$---where the
negative NNLO correction becomes stronger.
Note that for $M^{2}=1.5$~GeV$^2$ \cite{ABOP10}, the twist-six term
becomes negligible, while NNLO$_{\beta_{0}}$ still provides
suppression.
\item Inclusion of resonances: A finite-width Breit-Wigner form is
adopted in the spectral density in order to resolve the $\rho$ and
$\omega$ resonances \cite{Kho99,MS09}.
This entails a small enhancement of the transition form factor as
compared to the $\delta$-function ansatz.
\end{itemize}
Figure \ref{fig:1} illustrates the approximate composition of the
pion-photon transition form factor.

\section{Theoretical Predictions and Comparison with Data}
\label{sec:predictions}
\begin{figure*}
\centering
\centerline{\includegraphics[width=0.36\textwidth]{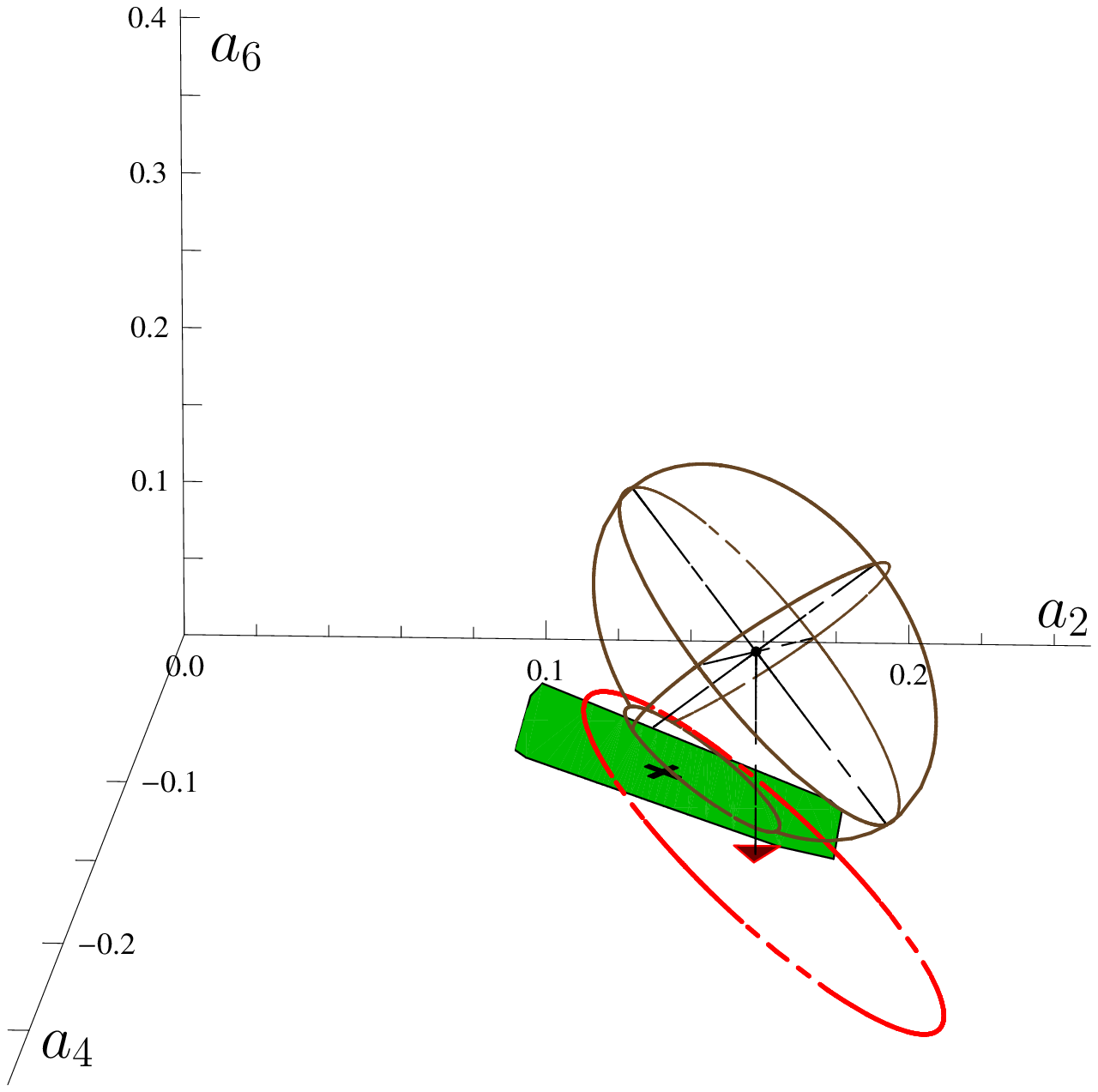}
~~~~~~~~~~~~\includegraphics[width=0.36\textwidth]{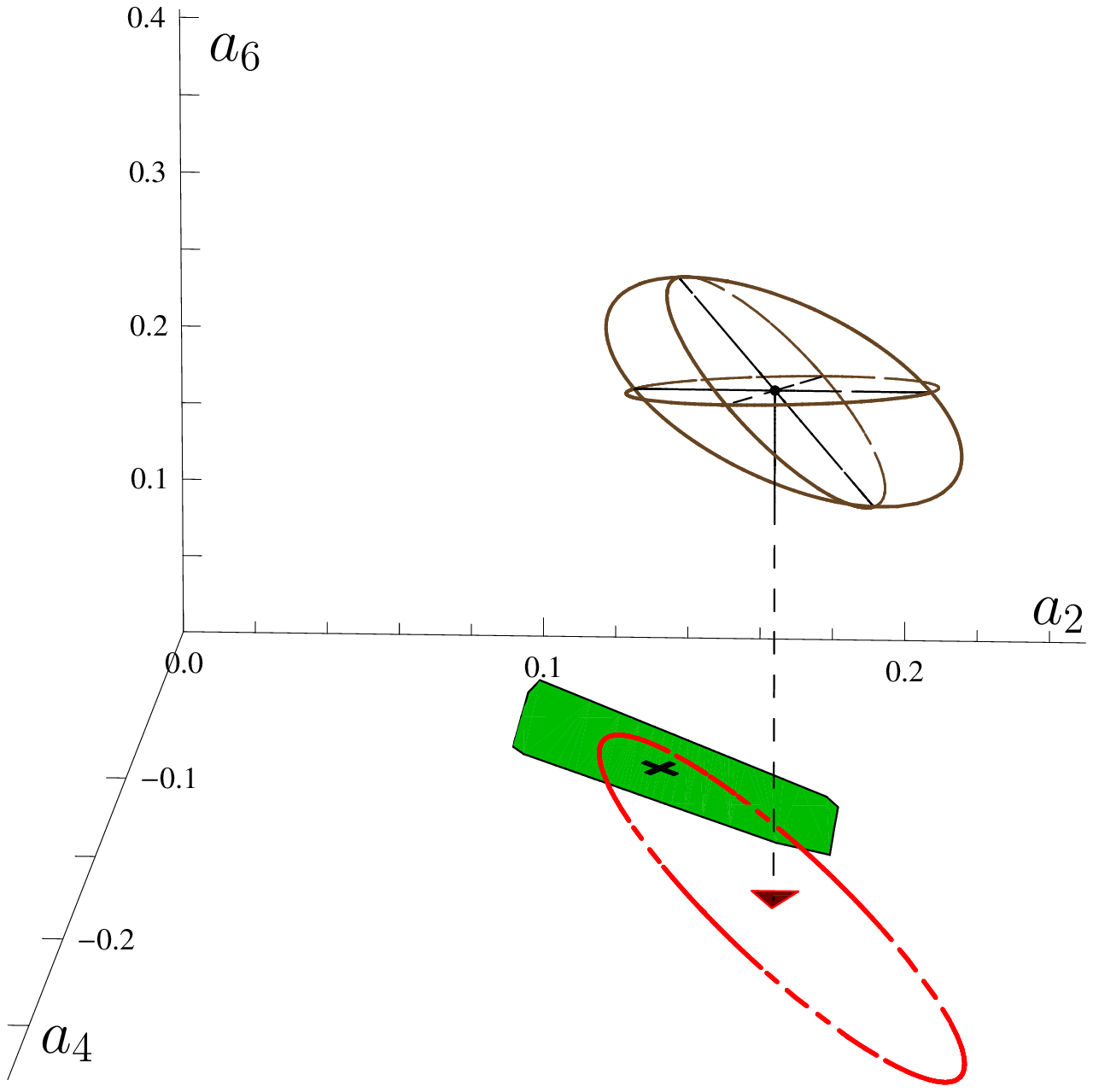}}
\vspace{20pt}
  \centerline{\includegraphics[width=0.40\textwidth]{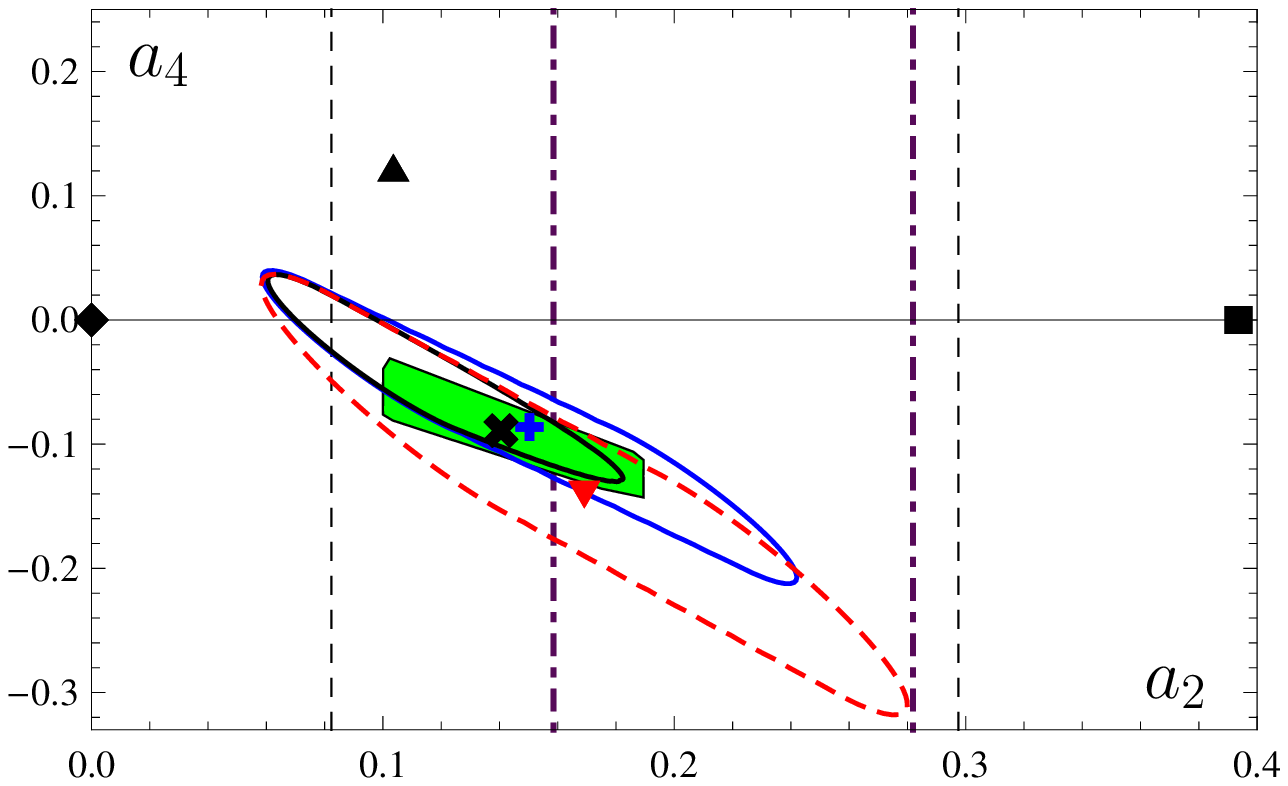}%
~~~~~~~~~~~~\includegraphics[width=0.40\textwidth]{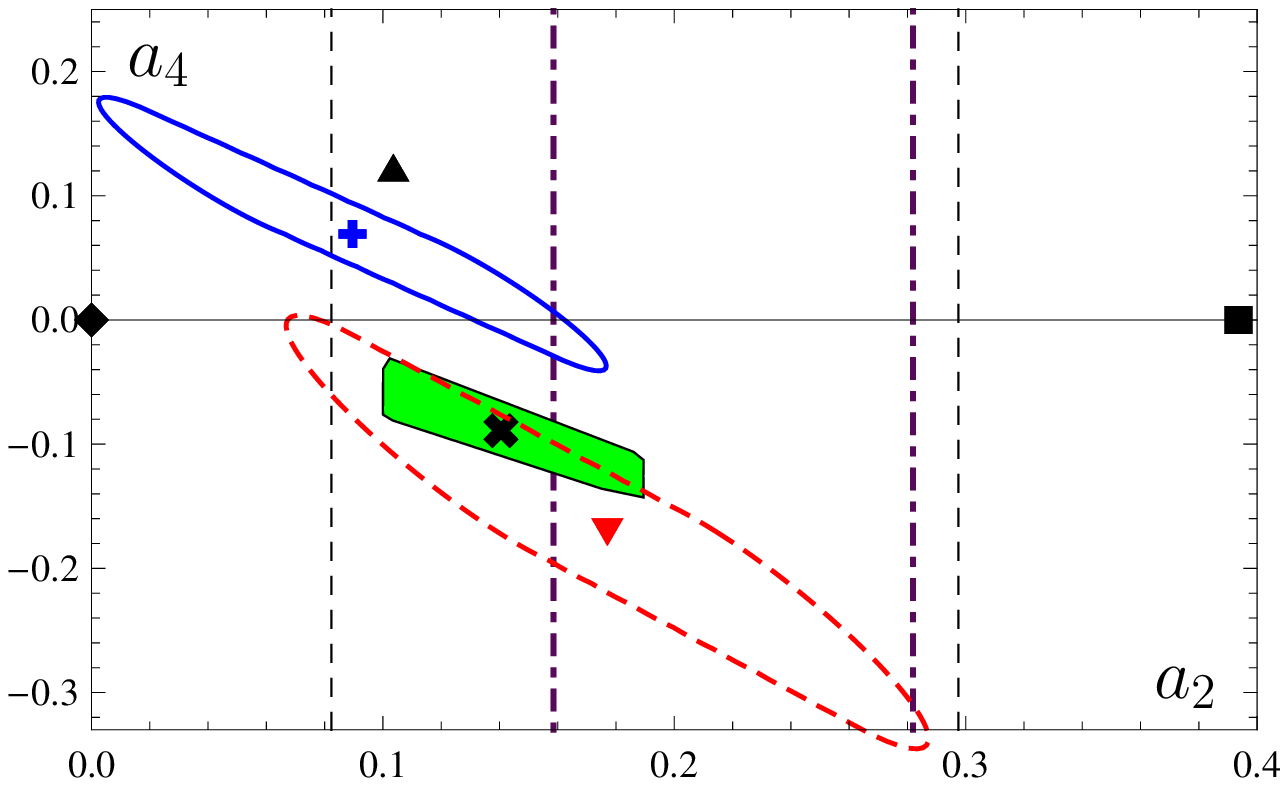}}
\caption{(color online).
Upper panels show 3-D graphics of $1\sigma$ error ellipsoids in
Gegenbauer space $(a_2, a_4, a_6)$.
Lower panels show $1\sigma$ error ellipses in the $(a_2, a_4)$ space
resulting from unifying $1\sigma$ ellipses pertaining to the values
of the twist-four coupling
$\delta^2=0.15, 0.19, 0.23$~GeV$^2$.
Left panels refer to the analysis of all data
\protect\cite{CELLO91,CLEO98,BaBar09} in the range $[1\div 9]$~GeV$^2$,
whereas the right panels give the analogous results
for the region $[1\div 40]$~GeV$^2$.
}
\label{fig:2}
\end{figure*}

We now have all the elements for the data analysis.
In order to get the most output from the data for the pion-photon
transition form factor \cite{CELLO91,CLEO98,BaBar09} and achieve the
best insight about the particular role and relative strength of the
first three Gegenbauer coefficients $a_2, a_4, a_6$, we divide the
data into two sets:
(a) $[1\div 9]$~GeV$^2$---termed `CLEO regime' and
(b) $[1\div 40]$~GeV$^2$---`entire range'.
The results of the statistical 3-D analysis of these data sets are
displayed in Fig.\ \ref{fig:2}---upper part---in the left and
right panel, respectively, using the central value
$\delta^2=0.19$~GeV$^2$ of the twist-four coupling.
Both panels show the corresponding $1\sigma$ ellipsoids which are
characterized by their principal axes and ellipses.
The projection of the $1\sigma$ ellipsoid on the plane $(a_2,a_4)$
is represented, in both panels, by the larger ellipse enclosed by a
solid line in red color.
The smaller enclosed ellipse in the left panel denotes the cross
section of the ellipsoid with the $(a_2,a_4)$ plane, whereas the
shaded (green) rectangle shows the region in the $(a_2,a_4)$ plane
allowed by nonlocal QCD sum rules \cite{BMS01}, with the point in
the center (marked by {\footnotesize\ding{54}}) denoting the
Bakulev-Mikhailov-Stefanis (BMS) pion DA \cite{BMS01}.
The profile of this DA exhibits a double-humped structure with
suppressed endpoints $x=0,1$ due to the use of nonlocal condensates
\cite{BMS01}---see also \cite{SSK99-00} for a discussion of the
endpoint region of the pion DA.
All results are shown at the scale $\mu_{\rm SY}^2=(2.4~{\rm GeV})^2$
after NLO evolution.
Performing the analysis with the second data set (upper right panel),
entails a sizeable coefficient $a_6$, so that the $1\sigma$ ellipsoid
is lifted off the plane $(a_2,a_4)$, while its projection remains
almost the same as in the left panel.
Remarkably, it still overlaps with the shaded rectangle containing
the BMS model and the constraints from nonlocal QCD sum rules
\cite{BMS01}.

Attempting to understand more quantitatively how the twist-four
contribution affects our results, we concentrate on the $(a_2,a_4)$
plane and present our findings in the lower part of Fig.\ \ref{fig:2}
in terms of $1\sigma$ ellipses.
These are calculated by varying the $\delta^2$ coupling from 0.15 to
0.19 to 0.23~GeV$^2$ and then merging the obtained $1\sigma$ error
ellipses together to form the distorted ellipses shown in this
graphics.
Left panel (`CLEO regime'): The largest ellipse---dashed (red)
line---results from combining the projections on the plane
$(a_2,a_4)$ of the 3-D data analysis discussed before.
The smaller ellipse (solid blue line) shows the outcome of a 2-D
analysis by means of $a_2$ and $a_4$.
Its middle point with the coordinates $(0.15,-0.09)$
and $\chi^2_{\rm ndf} \approx 0.5$
almost coincides with the center of the rectangle from \cite{BMS01}.
Finally, the smallest ellipse (thick solid line), entirely enclosed
by the previous one, is obtained by unifying the intersections with the
$(a_2,a_4)$ plane of all ellipsoids generated by the variation around
the central value of $\delta^{2}=0.19$~GeV$^2$.
Right panel (`entire range'): Including into the statistical analysis
the high-$Q^2$ tail of the \babar data modifies this picture
completely.
Now the composed error ellipse resulting from the 2-D analysis
(solid blue line) moves out of the region of the negative values of
$a_4$---characteristic of the BMS models \cite{BMS01}---entering
the positive domain.
This entails a significantly worse value
$\chi^2_{\rm ndf}\approx 2$,
relative to the value $\chi^2_{\rm ndf}\approx 0.5$
for the `CLEO-regime' data set.
On the other hand, the unified $1\sigma$ error ellipse of the 3-D
projections on the $(a_2,a_4)$ plane (larger dashed red ellipse)
keeps its position fixed still enclosing the major part of the
$a_2$, $a_4$ values obtained from nonlocal QCD sum rules
(shaded rectangle in green color).
Confronting these results with lattice estimates reveals that
the 3-D projected error ellipse lies almost entirely within the
boundaries from \cite{Lat06} (dashed vertical lines), even
intersecting for the larger values of $a_2$ with the narrower
interval determined in \cite{Lat10} (dashed-dotted vertical lines).
In contrast to the left panel, the ellipse of the 2-D analysis
for the entire set of data only poorly complies with the small
$a_2$ window of \cite{Lat10}, while partly overlapping with the low
end of the $a_2$ range determined in \cite{Lat06}.
For comparison, some characteristic pion DAs are also shown in
Fig.\ \ref{fig:2}:
asymptotic DA ({\footnotesize\ding{117}});
Chernyak-Zhitnitsky (CZ) model \cite{CZ84} ({\footnotesize\ding{110}});
projection of Model III from \cite{ABOP10}
({\footnotesize\ding{115}}).

\begin{figure*}
\centering
\centerline{\includegraphics[width=0.46\textwidth]{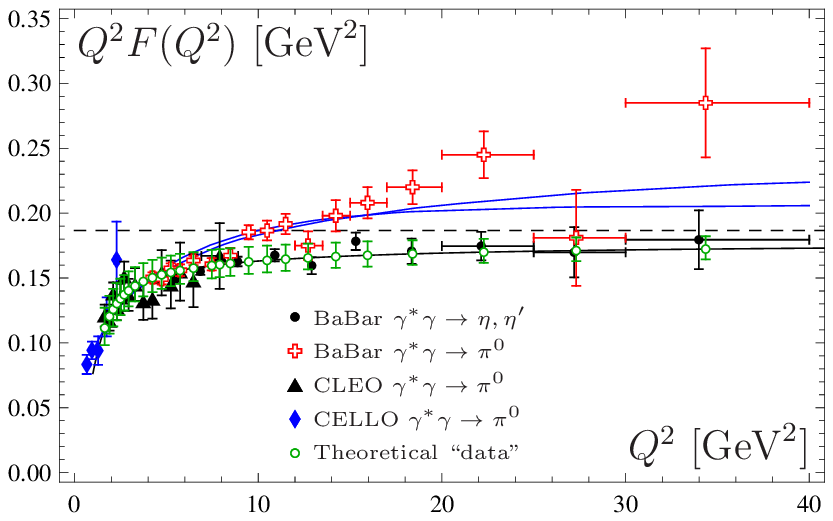}
  ~~~~~~~~~\includegraphics[width=0.46\textwidth]{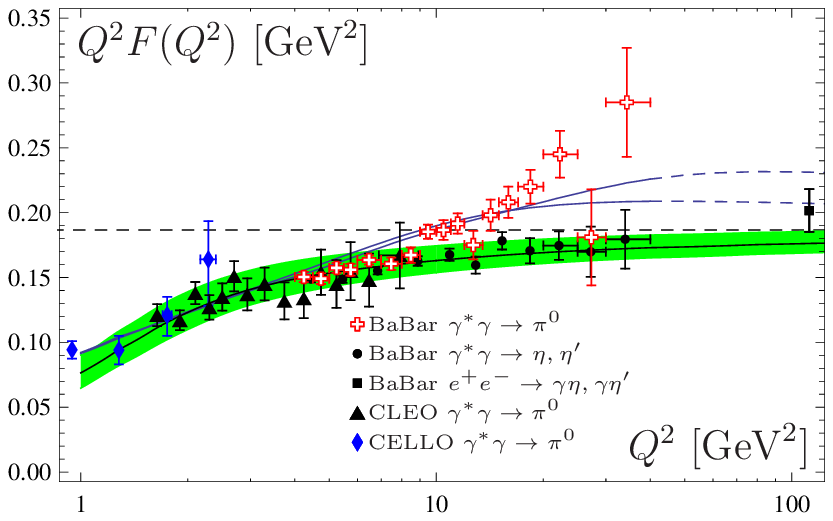}}
\caption{(color online).
$Q^2F^{\gamma^{*}\gamma\pi^{0}}(Q^2)$ as a function of $Q^2$.
Left panel shows ``benchmark theoretical data'' (open circles) that
include systematic uncertainties from different sources, explained in
the text, in comparison with real experimental data
\protect\cite{CELLO91,CLEO98,BaBar09,BaBar11-BMS}.
Right panel serves to show the large $Q^2$ form-factor behavior
using a logarithmic plot.
The shaded (green) strip contains the results for the BMS bunch of
$\pi$ DAs, with the BMS model \protect\cite{BMS01} being represented by
a solid line.
Also shown are the predictions of Agaev {\it et al.}
\protect\cite{ABOP10} for their models I and III, extrapolated in
terms of dashed (blue) lines to the remote timelike point 112~GeV$^2$
from \protect\cite{Aub2006}.}
\label{fig:3}
\end{figure*}

Figure \ref{fig:3} shows the calculated pion-photon transition form
factor \cite{BMPS11} in comparison with various sets of data,
specified in the caption.
The graphics in the left panel displays the theoretical results in the
form of ``data'' at the same momentum-transfer values as the
experimental data and with theoretical uncertainties included as error
bars---see Table I in the first work of Ref.\ \cite{BMPS11} for the
numerics.
The right panel serves to effect the large-$Q^2$ behavior of the
scaled form factor---therefore, a logarithmic scale.
The shaded green strip embodies the predictions obtained in
\cite{BMPS11}, as described above.
One observes:
(i) The shaded strip fits the CELLO, CLEO and the `CLEO regime' set of
the \babar data very precisely (overall $\chi^2 < 1$).
(ii) The prediction approaches asymptotically the limit $\sqrt{2}f_\pi$
from below and agrees very well with the \babar data for the
$\gamma^*\gamma\eta(\eta')$ transition form factor, being in conflict
with the steep rise of the \babar data for
$Q^2F^{\gamma^*\gamma\pi^0}(Q^2)$ above 9~GeV$^2$.
(iii) The LCSR predictions of \cite{ABOP10} (single solid blue lines)
are not really reproducing the behavior of the \babar data, despite
the opposite claims by the authors and the use of pion DAs with a high
number of harmonics.
In fact, the reinforcement of the form factor can only be achieved
by an enhancement of the endpoint region $x=0,1$ of the pion DA,
with the best agreement being provided by the flattop DA which
includes all $x$ values in an equal amount, inevitably entailing
a worse fit to the data in the `CLEO regime' and below
(see \cite{BMPS11} for details).

\section{Conclusions}
\label{sec:concl}

To conclude, we cannot \emph{predict} the rise of the scaled
pion-photon transition form factor, observed by the \babar
Collaboration, using the standard scheme of QCD.
Nor can we \emph{explain} it in hindsight within the same
context---despite opposite claims in \cite{ABOP10} and elsewhere
\cite{Kro10}.
Therefore, the \babar data are unpredictable in QCD, but they are not
unexplainable---in particular contexts which, however, are contingent
on assumptions that can hardly be accommodated within the standard
scheme of QCD.
Moreover, there is a dichotomy of the pseudoscalar meson sector into
two very disparate groups, one obeying the QCD asymptotics, the other
not, depending on the particular shape of the pion DA.
Indeed, for BMS-like models (double-humped and endpoint-suppressed)
\cite{BMS01}, the predictions completely agree with the CELLO
\cite{CELLO91} and \emph{all} data in the `CLEO region'
\cite{CLEO98,BaBar09} for the pion-photon transition form factor,
as well as with the \babar data \cite{BaBar11-BMS} for the
$\eta(\eta^\prime)$-photon transition form factor, while being in
strong disagreement with the \babar data \cite{BaBar09} for the
$\pi^0$-photon form factor above 9~GeV$^2$.
On the other hand, using a flatlike DA \cite{Rad09,Pol09,MPS10}
(which describes an unrealistic pointlike pion) reproduces the
$\gamma^*\gamma\pi^0$ \babar data but fails to comply with those on
the $\eta(\eta^\prime)$-photon transition leading to predictions
above the experimental data \cite{BMPS11} (see also
\cite{BCT11,Roberts10-11,WY11}).
These findings indicate a possible strong violation of the isospin
symmetry in the pseudoscalar meson sector ($\pi^0$ and $\eta_8$) that
has not been observed in other measurements so far.
Therefore, the reproducibility of the \babar results for the
pion-photon transition form factor by other experiments
(e.g., by Belle) is of paramount importance.

\begin{acknowledgements}
I am thankful to A.P.~Bakulev, S.V.~Mikhailov, and A.V.~Pimikov for a
fruitful collaboration within the Heisenberg-Landau
Program (Grant 2011), and to the DAAD for a travel grant.
\end{acknowledgements}



\end{document}